\documentclass[pra, showpacs,longbibliography,reprint]{revtex4-2} 

\usepackage{amsmath, amssymb, stmaryrd, braket, mathrsfs,mathtools,amsfonts}
\usepackage{bbold}
\usepackage{tcolorbox, accents}
\usepackage[colorlinks=true,citecolor=cyan,linkcolor=blue]{hyperref}
\usepackage{url}
\usepackage[toc,page]{appendix}
\usepackage{longtable,booktabs}
\usepackage{hyperref}
\usepackage{color}
\usepackage{float}
\usepackage{graphicx}
\usepackage{epsfig}
\usepackage{dcolumn}
\usepackage{bm}
\usepackage{multirow}
\usepackage[all]{xy}
\usepackage{pbox}
\usepackage{lipsum}
\usepackage{verbatim}
\usepackage{braket}
\usepackage{dsfont}
\usepackage{array}
\usepackage{makecell}
\usepackage{tabularx}
\usepackage{isotope}

\usepackage{fancyhdr}
\usepackage{amsthm}
\usepackage{amsfonts}
\definecolor{ceruleanblue}{rgb}{0.16, 0.32, 0.75}

\begin{document}
\title{Error recovery protocols within metastable Decoherence-Free Subspaces}
\author{Thomas Botzung}
\affiliation{CESQ/ISIS (UMR 7006), CNRS and Universit\'{e} de Strasbourg, 67000 Strasbourg, France
}
\author{Eliana Fiorelli}
\affiliation{Institute for Cross-Disciplinary Physics and Complex Systems (IFISC) UIB-CSIC,
Campus Universitat Illes Balears, 07122 Palma de Mallorca, Spain}

\begin{abstract}
    Open quantum systems governed by quantum master equations can exhibit quantum metastability, where decoherence-free subspaces (DFS) remain approximately invariant for long transient times before relaxing to a unique steady state. In this work, we explore the use of such metastable DFS as code spaces for passive quantum error correction. We focus on two representative models: a two-qubit system under collective dissipation, and a nonlinear driven-dissipative Kerr resonator. After characterizing the parameter regimes that support metastability, we introduce and analyze a protocol for error recovery during the metastable dynamics. Using spectral properties of the Liouvillian, we characterize which types of errors can be possibly autonomously reversed. In particular, we show that in the qubit model, the state affected by either bit-flip error or spontaneous emission can be recovered up to a certain measure. Instead, phase-flip errors would require further strategies. For the bosonic system, we show that dephasing-induced errors on cat states can be partially recovered, with a trade-off between fidelity and recovery time. These findings highlight the limitations and capabilities of metastable DFS as a transient resource for error correction.
\end{abstract}
    
\maketitle

\onecolumngrid

\section{Introduction}

The reliable storage and manipulation of quantum information face major challenges due to the accumulation of small errors and the destructive effects of decoherence arising from uncontrolled interactions with the environment~\cite{Shor1995, Steane1996}. Overcoming these issues is essential for the realization of scalable quantum computation. The predominant approach to quantum error correction (QEC) tackles this challenge by redundantly encoding logical information across many physical qubits and actively detecting and correcting errors, forming the basis of most current quantum computing architectures~\cite{Roffe2019Jul, Georgescu2020Oct}.

An alternative paradigm, known as passive quantum error correction~\cite{BookCambridge2013}, offers a fundamentally different route. Instead of actively detecting and correcting errors, this approach exploits the intrinsic dynamics of the system—often including its coupling to an environment—to autonomously stabilize encoded information. In such systems, dissipative processes associated with specifically engineered Hamiltonians can continuously steer the system back into the code space, thereby suppressing the accumulation of errors. For instance, the thermal bath associated with certain Hamiltonians can naturally induce dissipative dynamics that actively correct thermal errors~\cite{Bacon2006, Alicki2010, Chesi2010Aug, Yoshida2011Oct, Haah2011Apr, Pastawski2011Jan, Bravyi2013Nov, Bombin2013May, Terhal2015Apr, Bombin2015Sep, Breuckmann2016, Williamson2016Oct, Brown2016Nov, Bridgeman2024Apr, Herold2015Oct, Herold2017Jun, Ruiz2023Apr}. A notable example is the four-dimensional (4D) toric code~\cite{Dennis2002, Alicki2010}, which exhibits a self-correcting behavior: when the system is coupled to a thermal bath at a temperature below a critical threshold (in the thermodynamic limit), quantum information encoded in its ground state remains recoverable at any finite time. In this regime, thermal errors are counteracted by the system's intrinsic dynamics, and the encoded information is protected without the need for active error correction.

Much of the research in this area has focused on open quantum systems that exhibit multiple steady states, as these can retain memory of initial conditions~\cite{Buca2012Jul, Albert2014Feb, Albert2016Nov, Buca2019Apr, Roberts2020Apr, Macieszczak2016Jun, Chiacchio2019May, Dutta2020Dec, Cian2019Aug, vanCaspel2018May, Gau2020Oct, Zhang2020May, Santos2020Nov}. Recent work has established connections between these steady-state degeneracies and strong symmetries in the GKS–Lindblad dynamics~\cite{LieuEtAl20}, as well as connections between passive error correction and dissipative phase transitions: both hinge on the emergence of a robust, degenerate steady-state manifold, often with topological features~\cite{Liu2024}. However, while finite-size systems can display this kind of degeneracy, it tends to be fragile under arbitrary local perturbations.  Yet, identifying general classes of open systems that support such robust steady-state qubit structures remains a challenging task.

To face this challenge, metastable manifolds appear to be promising candidates for quantum information encoding. They have been shown to offer surprising advantages, for instance by converting decays out of the qubit subspace into erasure channels ~\cite{Ma2023Oct}. Metastable qubits have been proposed and demonstrated in both neutral atom~\cite{Wu2022Aug, Chen2022May} and ion trap platforms ~\cite{Allcock2021Nov, Campbell2020Aug}, notably in systems using $^{171}$Yb ions~\cite{Yang2022Sep}. More recently, metastable manifolds have been explored and formalized to host instances of quantum associative memories, which display the ability to restore corrupted inputs, and can be thought of as tools to preserve information in quantum memories \cite{labay2022memory, LabayMora25jun}.

These developments suggest that metastability, far from being a hindrance, may offer a useful operational regime for protecting quantum information—especially in open quantum systems where strong symmetries are only approximate and full steady-state degeneracy is not achievable. In such cases, the logical qubit structure may not be preserved indefinitely but can persist over a long-lived intermediate timescale before ultimately decaying into a trivial steady state. That is, the entire qubit manifold becomes metastable, rather than forming part of the system's asymptotic behavior \cite{MacieszczakEtAl_PRR_21, Macieszczak2016Jun, Brown2016Nov}.

In this work, we present two examples of this phenomenology and analyze the conditions under which error correction remains feasible during these metastable time windows. Our aim is to identify whether and how passive protection mechanisms can still operate effectively, even in the absence of truly degenerate steady states.

This paper is organized as follows: in Section \ref{Sec:metastability} we summarize the key concepts and tools for dealing with metastability in open quantum systems, focusing in particular on quantum metastability. In Section \ref{s:two qubit} we present a two-qubit model, discussing its metastable properties and presenting the protocol for error recovery. Section \ref{s: boson model} deals with similar concepts applied to a bosonic system. Finally, we draw our conclusions and perspective research in Section V.

\section{Metastability in open quantum systems}
\label{Sec:metastability}

In this Section, we introduce the type of open quantum systems we will deal with, and we summarize the main concepts for describing and employing quantum metastability. For a detailed treatment on metastability in open quantum systems, we refer to \cite{MacieszczakEtAl_PRR_21, Macieszczak2016Jun} and related works. 

We consider an open quantum system evolving under a Markovian dynamics, the time evolution of the state $\rho(t)$ being governed by the Gorini–Kossakowski–Sudarshan and Lindblad (GKS-Lindblad) equation
\begin{equation}
    \dot{\rho} = \mathcal{L}(\rho) = -i[H,\rho] + \sum_{j}J_j \rho J_j^{\dagger} - \frac{1}{2}\lbrace J_j^{\dagger}J_j, \rho\rbrace \, ,
\end{equation}
where $H$ is the Hamiltonian of the system, $J_k$ are the so-called jump operators \cite{Lindblad76, GorinKSi76}, and we omitted the time dependency for the sake of a lighter notation. As the Liouvillian $\mathcal{L}$ is a linear operator, one can express the above dynamics also in terms of the right eigenmatrices of the Liouvillian, $R_k$, with eigenvalues $\lambda_k \in \mathbb{C}$, that we will denote as $\lambda_k = \lambda_k^{R} + i \lambda_k^{I}$. For the sake of simplicity, we take the eigenvalues ordered by decreasing real part, $\lambda_1^R \geq \lambda_2^R \geq ... \geq \lambda_k^R \geq ... \, $. Moreover, assuming that the dynamics admits a unique steady state, $R_1 \equiv \rho_{ss}$, it corresponds to the zero eigenvalues $\lambda_1^R = 0$. The rest of the real part of the spectrum must be negative, $\lambda_k \leq 0$, $\forall k \geq 2$, as the GKS-Lindblad dynamics maps any physical state into a physical state, that is, it represents a completely positive and trace-preserving map. With the introduced notation, the state at the time $t$ can be written as
\begin{equation}
    \rho(t) = \rho_{ss} + \sum_k c_k e^{\lambda_k t}R_k \, ,
\end{equation}
the coefficient $c_k$ depending on the choice of the initial state $\rho(0)$ \footnote{Defining $L_k$ the left eigenmatrices of the Liouvillian operator, $c_k = \mathrm{Tr}[L_k \rho(0)$].}.

As thoroughly treated by Refs. \cite{Macieszczak2016Jun, MacieszczakEtAl_PRR_21}, metastability corresponds to a long-time regime in which the system appears stationary, before relaxing to the true stationary state, $\rho_{ss}$. For this to occur, there must be a large separation between the low-lying eigenvalues of the spectrum, taken in terms of the real part, and the rest of the spectrum. Let us assume that such a separation takes place between the $m$-th and the $(m + 1)$-th eigenvalue, $\lambda_m^{R}/\lambda_{m+1}^{R} \ll 1 $. Then the dynamics will appear stationary and read
\begin{equation}
    \rho(t) \approx \rho_{ss} + \sum_{k=2}^m c_k R_k \, ,
\end{equation}
during a time interval where the higher modes are negligible, i.e. $t \gg \tau' \equiv \frac{1}{|\lambda_{m+1}^{R}|}$, while the overlap of the initial state with the low-lying modes of the spectrum remains non-vanishing, which requires $t \ll \tau''\equiv \frac{1}{|\lambda_m^{R}|}$. Here, $\tau'$ and $\tau''$ are associated with a fast and a slow timescale, respectively. We denote the length of the time interval in which metastability takes place as
\begin{equation}\label{eq:meta_time interval}
    \Delta T_m = \tau''-\tau' \, .
\end{equation}

The minimal model where metastability can be analyzed corresponds to the case of two-phase metastability, which features $m = 2$. Here, the system exhibits a large separation between the second and third eigenvalues of the spectrum, $\lambda_2^R /\lambda_3^R \ll 1$.  Two distinct metastable
phases can be found, described by trace-$1$ matrices $\rho_1$, $\rho_2$. The latter are called extreme metastable states, and they are density operators \cite{MacieszczakEtAl_PRR_21, Macieszczak2016Jun} \footnote{Classical corrections apply if the metastable phases are not
sufficiently orthogonal \cite{Macieszczak2016Jun}}. They play a key role in describing the state of the system during the metastable time regime: After the fast timescale $\tau''$, and before the slow one, $\tau'$, the state of the system reads $\rho(t) \approx p_1 \rho_1 + p_2 \rho_2$, with $p_1, p_2$, behaving like probabilities depending on the overlap between the initial state and the phases. The metastability occurring for the case $m=2$ has been shown to share many
similarities to the one that occurs in simple classical systems. Indeed, when the metastable regime emerges separated in $m$ metastable phases, each of them is represented by a density operator, and the phenomenon is referred to as classical metastability . 

The case we will focus on in this work is more general and corresponds to the situation where coherences between phases are also metastable. This case requires $m>2$. In fact, we will deal with systems with $m= 4$, characterized by $\lambda_4^R/\lambda_5^R \ll 1$, and featuring two metastable states and two metastable coherences, a situation that corresponds to the existence of a metastable decoherence-free subspace (DFS) \cite{Lidar14}. Since this scenario supports a non-classical metastable structure, and more specifically a qubit metastable structure, this type of metastability is referred to as quantum metastability \cite{Macieszczak2016Jun, Brown2016Nov}. Here, identifying as $\ket{\psi_1}$, $\ket{\psi_2}$ the two metastable states, during the metastable time interval, $\tau' \ll t \ll \tau''$ the density matrix can be expressed as $\rho(t) \approx p_1(t) \ket{\psi_1}\bra{\psi_1} + p_2(t)\ket{\psi_2}\bra{\psi_2} + z(t) \ket{\psi_1}\bra{\psi_2}  + z^*(t)\ket{\psi_2}\bra{\psi_1}$, with $p_1$, $p_2$ real-valued probabilities, and $z(t )$ a complex number, with $|z|^2 \leq p_1 p_2$ \cite{Brown2016Nov}. Although we will focus on metastable DFS, it is worth mentioning that, in general, there can also be metastable noiseless subsystems \cite{MacieszczakEtAl_PRR_21}.

By exploiting the description and the tools introduced in this Section, in the following we will consider two models, specifying their Hamiltonian and jump operators. In both cases, we will focus on the parameter regime supporting quantum metastability, the latter corresponding, as in the above description, to a qubit structure. We will analyze how the presence of the quantum metastable time regime can effectively protect the system from errors, modeled in terms of jump operators. In the analysis performed for both the models of Section \ref{s:two qubit}, \ref{s: boson model}, our simulation code for the Liouvillian spectrum and state evolution was performed using the QuTip package in Python \cite{johansson2012qutip}. 

\begin{figure*}[htb]
    \centering
    \includegraphics[width=1\linewidth]{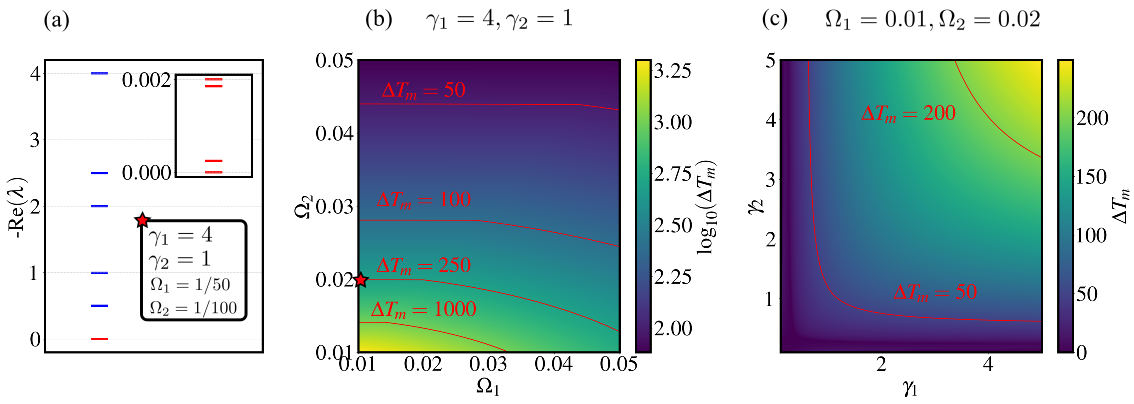}
  \caption{ Quantum metastability of  the two-qubit model subjected to collective jump operators. Panel (a) shows the first $9$ elements in the spectrum of the Liouvillian, in terms of their real part, the inset considering the first four eigenvalues. We set $\gamma_1=4, \gamma_2=1, \Omega_1=1/50, \Omega_2=1/100$. (b) Metastable time interval $\Delta T_m$,  in a logarithmic scale, upon varying the driving field  $\Omega_1$ and $\Omega_2$, with $\gamma_1=4, \gamma_2=1$. The (red) lines highlight from the the right-upper corner to the left-bottom one the following metastable time interval $ \Delta T_m = \left\lbrace 50, 100, 250, 1000 \right\rbrace $. (c) We set $\Omega_1=1/50, \Omega_2=1/100$ and display the metastable time window as the collective dissipation $\gamma_1$ and $\gamma_2$ are varied, with the red, continuous lines representing $\Delta T_m = \left\lbrace 50, 200\right\rbrace$.}
   \label{fig: transient2qubits}
\end{figure*}

\section{A two-qubit model with quantum metastability}
\label{s:two qubit}

In this section, we consider a two-qubit model that has been introduced in \cite{Macieszczak2016Jun}. The system is evolved by means of a GKS-Lindblad equation with a Hamiltonian reading
\begin{eqnarray}\label{e:twoq Hamilt}
    & H = \Omega_1\sigma_1^x + \Omega_2 \sigma_2^x  \, ,
\end{eqnarray}
and it features collective dissipation in terms of the following jump operator,
\begin{equation}\label{e:twoq collective jump}
     J = \sqrt{\gamma_1}n_1 \sigma_2^- + \sqrt{\gamma_2}(1-n_1)\sigma_2^+ \, .
\end{equation}
Here, $\sigma_i^{\alpha}$, $i=1,2$, $\alpha=x,y,z$ are the Pauli operators, $\sigma_i^{\pm} = \frac{1}{2}(\sigma_i^x \pm \sigma_i^y)$, and $n_i = \frac{1}{2}(1+\sigma_i^z)$. The model is characterized by a jump operator that annihilates the normalized superposition $a\ket{01} + b \ket{10}$, $\forall a,b \in \mathbf{C}$, $|a|^2+|b|^2 =1$. As a consequence, for sufficiently small values of the Hamiltonian parameters with respect to the dissipation rate, $\Omega_i \ll \gamma_j$, $\forall i=1,2$, $\forall j =1,2$, there exists a separation between the first $4$ eigenvalues of the corresponding Liouvillian, and the rest of the spectrum, $\lambda_4^R/\lambda_5^R \ll 1$. The manifold spanned by $\ket{01}$, $\ket{10}$ behaves as a metastable DFS. That is, for a relatively long time interval, all the states belonging to such a manifold are protected from both dissipation and decoherence, which gives rise to quantum metastability. It is worth mentioning that a similar model has been analyzed in Ref.~\cite{Brown2016Nov} as a quantum reset model, by considering two separated jump operators, the first with rate $\gamma_1$ and the second with rate $\gamma_2$. In this situation, the model features two reset points, $\ket{10}$, $\ket{01}$. Even in this case, the manifold spanned by the two reset-point states, undergoes quantum metastability in given parameter regime. 

The metastability regime of the model is characterized in Fig.~\ref{fig: transient2qubits}. Panel (a) shows the first $9$ eigenvalues of the spectrum of the Liouvillian given in terms of Eqs.\eqref{e:twoq Hamilt}, \eqref{e:twoq collective jump}, the inset showing the first $4$ eigenvalues. Panels (b), (c) characterize the metastability time interval when varying the parameters of the model. As shown by Ref. \cite{Macieszczak2016Jun}, the regime with $ \Omega_i \ll \gamma_j$ supports a metastable DFS. In panel (b), we consistently observe that reducing $\Omega_{1,2}$ with respect to $\gamma_{1,2}$ extends the quantum metastable regime. Likewise, in panel (c) we highlight that increasing the rates $\gamma_1$, $\gamma_2$, while fixing $\Omega_i$ also leads to a longer metastable regime.

\label{s:metadfs}
\subsection{Error recovery protocol}

\begin{figure*}[p]
    \centering
    \includegraphics[width=0.95\linewidth]{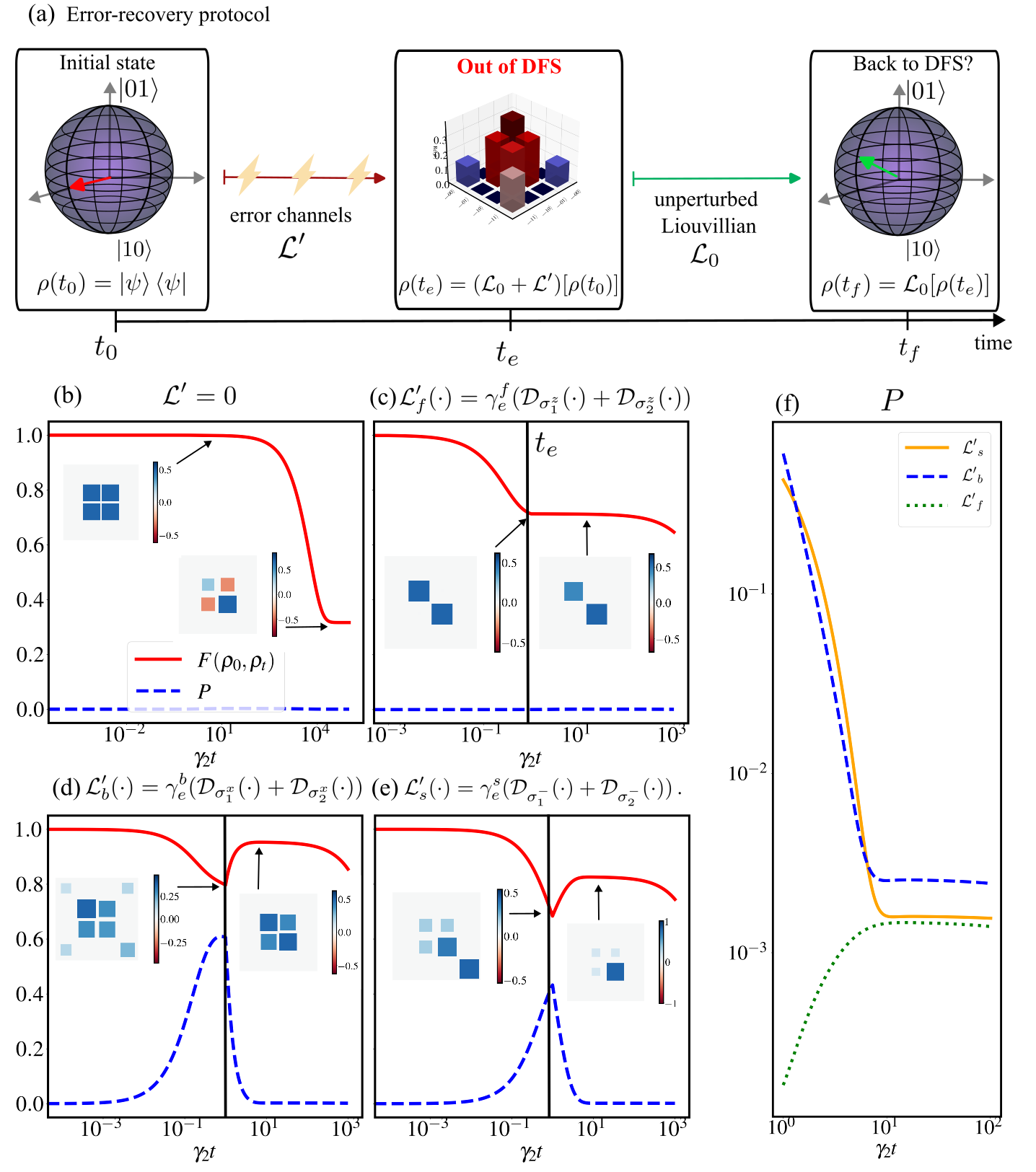}
   \caption{(a) Error-recovery protocol. The state is prepared within the metastable DFS of Liouvillian $\mathcal{L}_0$, evolved with the addition of the error channel, $\mathcal{L}'$, up to the time $t_e$, and let evolve with the unperturbed Liouvillian $\mathcal{L}_0$ up to a final time $t_f$, such that $\Delta t_e \ll \Delta t_f < \Delta T_m$. Evolution through quantum metastability for an unfaulty preparation (b), with a phase flip error model (c), a bit flip error model (d), and spontaneous emission (e). In panel (b) the time evolution is displayed for a longer time interval than the metastable transient time, $\Delta T_m \approx 10^3$, while panels (c), (d), (e) show the evolution within the metastable transient time, the vertical line highlighting the error time $t_e$ after which the error channel is turned off. The error time is $\gamma_2t_e=1$. Lines shows the fidelity (full lines) between the initial state, $\rho_0$, and the evolved state, $\rho_t$, and the expectation value of the projector outside the DFS (dashed lines) as a function of time. We fix $\gamma_2=1$, $\gamma_1=4$, $\Omega_1=0.02$, $\Omega_2=0.01$, and $\gamma_e = 1$. The initial state is chosen as a state belonging to the metastable DFS ($\psi_0 = (\ket{01} + \ket{10})/\sqrt{2}$). The insets show the state representation of the state evolved at the time $\gamma_2 t =1$, and $\gamma_2 t =10$. (f) Expectation value of the projector outside the DFS as a function of time, for $t \geq t_e$. Different line corresponds to different errors occurred in the time interval $[0,t_e]$, i.e. spontaneous emission (orange line), bit-flip error (blue, dashed line), phase-flip error (green dotted line).}
   \label{fig: 2qubitserrors}
\end{figure*}

After having discussed the two-qubit model, and in particular the regime where it exhibits a metastable DFS, we will now introduce a possible scheme of error recovery that exploits the presence of quantum metastability itself. To this end, we consider the Liouvillian specified in terms of Hamiltonian and jumps operator defined by Eqs. \eqref{e:twoq Hamilt}, \eqref{e:twoq collective jump}, and we refer to such a superoperator as ``unperturbed'' Liouvillian, denoting it by $\mathcal{L}_0$. We will then deal with the parameter regime in which the system exhibits quantum metastability in terms of the DFS spanned by the states ${\ket{01}, \ket{10} }$. Such a metastable DFS represents a code space that could eventually be employed in quantum protocols or algorithms \cite{LidarCW98}. Here, we are interested in understanding and characterizing how the system is robust with respect to error channels, described by an additional superoperator, $\mathcal{L}'$, acting for a limited amount of time and causing the system to possibly exit the metastable DFS. A similar type of error model, acting for a finite amount of time and characterized by fixed parameters, could represent, e.g.,  an error source while applying an imperfect gate. Before going ahead and specifying the details of the channel causing faults in the system state, we describe the protocol taken into account, which is schematically depicted in Fig.~\ref{fig: 2qubitserrors} (a). It consists in the following three steps:

\begin{itemize}
    \item[$1)$] Initialize the system at time $t_{0} = 0$ in a state, say $\ket{\psi}$, belonging to the metastable DFS of the unperturbed Liouvillian $\mathcal{L}_0$, $\rho(t_0) = \ket{\psi}\bra{{\psi}}$.
    \item[$2)$] Let an error occur in terms of an additional Liouvillian $\mathcal{L'}$, that acts from $t_{0}= 0$ to a final time $t_e$, that is, during a time interval $\Delta t_e $ smaller than the metastable time window, $\Delta t_e \ll \Delta T_{{m}}$, so that the initial state evolves into $\rho( t_e) = (\mathcal{L}_0 +\mathcal{L}')[\rho(t_0)]$.
    \item[$3)$] Let the dynamics ruled by the unperturbed Liouvillian $\mathcal{L}_0$, evolve the state $\rho( t_e)$ up to a final time, say $t_f$, $\rho(t_f) = \mathcal{L}_0[\rho( t_e)]$.
\end{itemize}
Within this procedure, we want to understand how the presence of a quantum metastable manifold can protect the system after some errors occurred. As such, we will focus on the fidelity between the initial state $\rho(t_0) \equiv \rho_0$ and the generic evolved state $\rho(t) \equiv \rho_t$, to quantify the (possible) evolution back to the initial state after the error. 
In order to proceed further, we define the types of error channel that we will consider. 

\subsubsection{Bit-flip, phase-flip error, and spontaneous emission}

The first error we consider induces phase-flip type of faults, and it is modeled in terms of the following jump operators
\begin{equation}
    J_{i, \gamma_e}^{f} = \sqrt{\gamma_e^f} \sigma_i^{z},
\end{equation}
for $i=1,2$, the corresponding error channel reading
\begin{equation}
    \mathcal{L}_f'(\cdot) = \gamma_e^f(\mathcal{D}_{\sigma_1^{z}}(\cdot) +\mathcal{D}_{\sigma_2^{z}}(\cdot) ) \, ,
\end{equation}
with $\mathcal{D}_O(\cdot) = O(\cdot)O^{\dagger} - \frac{1}{2}\lbrace O^{\dagger}O,\cdot \rbrace.$ Similarly, we take into account the bit-flip error, which is given in terms of the jump operator
\begin{equation}
    J_{i, \gamma_e}^{b} = \sqrt{\gamma_e^b} \sigma_i^{x},
\end{equation}
for $i=1,2$, corresponding to the following error channel
\begin{equation}
    \mathcal{L}_b'(\cdot) = \gamma_e^b(\mathcal{D}_{\sigma_1^{x}}(\cdot) +\mathcal{D}_{\sigma_2^{x}}(\cdot) ) \, .
\end{equation}
Finally, we will also deal with spontaneous emission, which can be modeled via the jump operator
\begin{equation}
    J_{i, \gamma_e}^{s} = \sqrt{\gamma_e^s} \sigma_i^{-},
\end{equation}
and corresponding error channel
\begin{equation}
    \mathcal{L}_s'(\cdot) = \gamma_e^s(\mathcal{D}_{\sigma_1^{-}}(\cdot) +\mathcal{D}_{\sigma_2^{-}}(\cdot) ) \, .
\end{equation}
In the following, we take the same error rate for all the error channels, $\gamma_e^s = \gamma_e^b = \gamma_e^f \equiv \gamma_e$.
We then apply the protocol described above and characterize it by analyzing $i)$ how the error channel affects the code space, $ii)$ how the error channel affects the state, and finally $iii)$ to what extent the presence of a metastable DFS is able to restore the state to the initial one, after the error occurred. As already mentioned, a key quantity that we consider is the fidelity between the initial state and the evolved state, which is defined as
\begin{equation}\label{e:fidelity}
    F(\rho_0, \rho_t) =  \sqrt{\mathrm{Tr}(\sqrt{\rho_0} \rho_t \sqrt{\rho_0})} \, .
\end{equation}
Secondly, we also consider the projector outside the code space, 
\begin{equation}\label{e:projector out}
    P = \mathbf{1} - \ket{01}\bra{01} - \ket{10}\bra{10} \, ,
\end{equation}
where $\mathbf{1}$ is the identity on the two-qubit Hilbert space. This operator allows us to quantify if and to which measure the error channel causes the state to exit the metastable DFS. 

The results for the different faulty protocols are illustrated in Fig.~\ref{fig: 2qubitserrors}, panels (c), (d), (e). For the sake of completeness, panel (b) shows the unperturbed evolution within the DFS and the eventual relaxation towards the steady state. Here, the state is prepared within the metastable DFS, and it is evolved by means of the unperturbed Liouvillian $\mathcal{L}_0$. The red line represents the time evolution of the fidelity between the initial state and the evolved one, $F(\rho_0, \rho_t)$. We can appreciate how the evolved state remains unperturbed for a time interval of the order $\Delta T_m$, eventually reaching the steady state $\rho_{ss}$. It is worth noticing that the latter is still a state that belongs to the metastable DFS. This is highlighted by the expectation value of the projector at time $t$, $\braket{P}_t$ (blue, dashed line), which remains indeed zero. 

The behavior of the state during the protocol that encompasses the errors 
is illustrated in panels (c), (d) and (e) of Fig.~\ref{fig: 2qubitserrors}, respectively. Let us first focus on the phase-flip error, illustrated in panel (b). We can see that the presence of the error channel suppresses the coherences between the states $\ket{01}, \ket{10}$ (see the inset that represents the density matrix), leading to a classical bit structure. Within a good approximation, the resulting mixed state still belongs to the metastable manifold spanned by $\ket{01}$ $\ket{10}$, and therefore the state does not exit the code space. Consistently with this, the expectation value $\braket{P}_t$ of the projector $P$ remains almost zero throughout the entire time evolution. However, the coherences between the states are completely lost, and so is the qubit structure. Therefore, in general, the presence of metastable DFS is not helpful in protecting the system from this type of errors.

A different result is obtained when considering the bit-flip error and spontaneous emission, as illustrated in panels (d), (e). Here, in both situations, the evolution in the presence of the error channel brings the state of the system outside the code space. Indeed, while the fidelity $F(\rho_0,\rho_t)$ decreases, the expectation value of the projector outside the code space, $\braket{P}_t$ increases, reaching the extrema at the time $\gamma_2 t_e$ when the error channel is switched off. Eventually, the presence of a quantum metastable manifold seems to be capable of leading the state back to the code space, in such a way that --up to a certain fidelity value-- the initial state can be restored. 

At variance with the phase-flip case, under either bit-flip error or spontaneous emission, the metastable DFS is approximately able to attract the state back into the DFS itself, reversing the effect of the occurred error. It is worth noticing that, although the error can be recovered as quantified in terms of the fidelity, there may remain some leakage outside the code space. To quantify such an occurrence, panel (f) shows the expectation value of the projector outside the metastable DFS, $\braket{P}_t$, after the time $t_e$ at which the error channels is switched off.  In the panel, we show the evolution of the projector after the spontaneous emission (orange line), bit-flip error (blue dashed line), and phase-flip error (dotted green line). We clearly see that in general its value is non-vanishing in all the cases, although being of order of magnitude slightly above $10^{-3}$. As such, while the state after error can be recovered within some threshold on the fidelity, the possibility to perform further operation in a qubit structure provided by the metastable DFS must be carefully analyzed.

The case treated in this section allows us to define the capabilities and limitations of the protocol described above in a simple yet meaningful example given by a two-qubit model. As we will employ later, a figure of merit to further quantify the recovery of the faulty state is the maximum value of the fidelity after an error occurred (i.e., at a later time than $t_e$). Accordingly, the time at which the fidelity assumes such a maximum value can be taken as recovery time, and identified with $t_r$. These quantities will be used to characterize the model that we take into account in the next section. 

\section{Two-photon driven-dissipative Kerr resonator}
\label{s: boson model}

The analysis performed in the previous section can be extended to other models, as already commented in Sec. \ref{s:metadfs}. In the following, we apply the analysis and the protocol previously introduced to a bosonic model. Before going ahead, we point out that bosonic codes are the subject of active research, the latter developing the main idea of encoding the information in bosonic modes \cite{Terhal2020jul, GillaudCM23}. Among the several approaches that have been followed in this context, 
one consists in encoding the information non-locally, by means of the phase space of harmonic oscillators \cite{GottesmanKP01, GillaudCM23, MirrahimiEtAl14, Gravina2023Jun}. One way to do so is to focus on the so-called cat encoding \cite{MirrahimiEtAl14, Gravina2023Jun, labaymora2025mar}. That is, denoting $\ket{\alpha}$ as a coherent state with complex amplitude $\alpha$, such that $a\ket{\alpha} = \alpha \ket{\alpha}$ for the annihilation operator $a$, the cat states are defined as
\begin{equation}
    \ket{C_{\alpha}^{\pm}} = \mathcal{N}_{\pm} (\ket{\alpha} \pm \ket{-\alpha}),  \quad \mathcal{N}_{\pm} = \frac{1}{\sqrt{2(1 \pm e^{-2|\alpha|^2})}} \, \, ,
\end{equation}
this giving the possibility to define the cat qubit states as
\begin{equation}
    \ket{0}_c = \frac{1}{\sqrt{2}}(\ket{C_{\alpha}^{+}} + \ket{C_{\alpha}^{-}}) \, , \quad \ket{1}_c = \frac{1}{\sqrt{2}}(\ket{C_{\alpha}^{+}} - \ket{C_{\alpha}^{-}}) \, .
\end{equation}

In the following, we will follow the above paradigm by focusing on a driven-dissipative Kerr resonator. The coherent part of the evolution, in a
frame rotating at the resonator frequency, reads
\begin{equation}\label{e:Kerr2pdriving}
    H = \frac{K}{2} (a^{\dagger}a)^2 + \frac{\Lambda_2}{2}[(a^{\dagger})^2+a^2] \, ,
\end{equation}
where we also introduced the creation operator $a^{\dagger}$, which, together with the annihilation one, satisfies the commutation relation $[a, a^{\dagger}] = 1$. The parameter $K$ is the Kerr non-linearity, and $\Lambda_2$ is the amplitude of the coherent two-photon driving. Let us also add a two-photon dissipation with jump operator
\begin{equation}\label{e:2phdissipation}
    J_2 = \sqrt{k_2}a^2 \, .
\end{equation}
The above model has been investigated in different regimes, with the aim of stabilizing a cat-state manifold. On the one hand, whenever $K = 0$, the system can easily be shown to be equivalent to a pure dissipative evolution with the jump operator $\tilde{J}= \sqrt{k_2}(a^2-\alpha^2)$, where $\alpha =\sqrt{\Lambda_2 / i\kappa_2} $ \cite{MirrahimiEtAl14}. The two coherent states $\ket{\pm \alpha}$ are thus dark states of the model, making the manifold spanned by the cat states $\ket{C_{\alpha}^{\pm}}$ a stationary $2$-dimensional space free from decoherence, i.e a DFS. Here, due to the non-local encoding of the cat qubit states, local noise inducing bit-flip errors can be efficiently suppressed, and the model turns out to be protected by dephasing errors when these occur for a shorter time than the one defined by the dissipative gap \cite{MirrahimiEtAl14}. On the other hand, a cat qubit can be engineered by means of a Hamiltonian, by exploiting Kerr non-linearity and two-photon driving \cite{PuriBB17}, when taking $k_2 = 0$ in the above model. In this case, the two coherent states $\ket{ \pm\beta} $ with $\beta= \pm i \sqrt{\Lambda_2/K}$ are eigenstates of the Hamiltonian. When considering the generic case of finite, non-vanishing $K$ and $k_2$, the stationary state is no longer a $2$-dimensional DFS. Nevertheless, the system displays a symmetry with respect to the parity operator, and it exhibits cat-bistability at stationarity. That is, the Liouvillian admits two steady-state solutions, corresponding to the two cat states $\ket{C_{\pm \alpha}}$, where now $\alpha =   i \sqrt{\frac{\Lambda_2}{K-i \kappa_2}} $ \cite{RobertsC20}. If the system is prepared in an initial state chosen in either one of the two symmetry sectors (i.e., even or odd), then it evolves to the corresponding cat state. For instance, the state $\ket{0}$ [$\ket{1}$] of the Fock basis is evolved into $\ket{C_{+\alpha}}$ [ $\ket{C_{-\alpha}}$]. However, any state that admits components in both sectors will end up in the mixture of the two cat states $a^2\ket{C_{+\alpha}}\bra{C_{+\alpha}} + b^2 \ket{C_{-\alpha}}\bra{C_{-\alpha}} $, with the population $a^2$, $b^2$, depending on initial conditions.

In the following, we are interested in adding also the effect of photon losses, which can be modeled in terms of the following jump operator
\begin{equation}\label{eq:single_photon_dissipation}
    J_1 = \sqrt{k_1}a \, .
\end{equation}
In the complete model, defined by Eqs.~\eqref{e:Kerr2pdriving}, \eqref{e:2phdissipation}, \eqref{eq:single_photon_dissipation},  
the density operator of the system evolves according to the Lindblad equation
\begin{equation}\label{eq:lindblad_kerr}
    \mathcal{L}_0[\rho] = -i[H,\rho] + \kappa_2 \mathcal{D}_{a^2}[\rho] + \kappa_1 \mathcal{D}_{a}[\rho],
\end{equation}
where $\mathcal{D}_{O}[\rho] = O \rho O^{\dagger} - \frac{1}{2} \lbrace O^{\dagger}O, \rho \rbrace$, and the Hamiltonian is defined by Eq.~\eqref{e:Kerr2pdriving}. This model has been analyzed by employing different techniques \cite{WolinskyC88, MirrahimiEtAl14, BartoloEtAl16, MingantiEtAl_nature_16, RobertsC20}, and, moreover, it has been experimentally realized in superconducting circuits \cite{LeghtasEtAl15}. 
Importantly, the presence of single-photon loss does not preserve the cat-qubit structure, and the system asymptotically evolves towards a unique steady state \cite{MingantiEtAl_nature_16, RobertsC20}. The latter can be diagonalized in terms of two catlike states for a broad range of parameters, including the ones compatible with the experimental realization $K \sim \kappa_2$, $\Lambda_2 \gtrsim \kappa_2 $, $\kappa_1 < \kappa_2$. However, and as pointed out in Ref. \cite{MingantiEtAl_nature_16}, the evolution towards such a stationary state strongly depends on the initial conditions. Moreover, there is an entire parameter regime where the evolution towards the steady state is characterized by metastable plateaus. In the next section, we will focus on such an occurrence.

\subsection{Quantum metastable regime}

In certain parameter regimes, the model defined by Eq.~\eqref{eq:lindblad_kerr} features quantum metastability. This is highlighted by a separation in the Liouvillian spectrum between the $4-$th and $5-$th eigenvalues, $\lambda_4^R/\lambda_5^R \ll 1$, as described in Section \ref{Sec:metastability}. By focusing on a parameter regime compatible with experimental realizations \cite{LeghtasEtAl15}, $  K \lesssim \kappa_2$, $\Lambda_2 \gtrsim  \kappa_2$, $0<\kappa_1<\kappa_2  $, within a good approximation \cite{MingantiEtAl_nature_16, Roberts2020Apr}, the metastable DFS is spanned by the cat states
\begin{equation}
    \ket{C_{\alpha, \pm}} = \frac{1}{\mathcal{N}}(\ket{+\alpha} \pm \ket{-\alpha} ) \, ,
\end{equation}
$\ket{\alpha}$ being the coherent states with $\alpha = i \sqrt{\frac{\Lambda_2}{(K-i\kappa_2)}}$. In other words, one can find a relatively long time interval where the dynamics is confined in the manifold spanned by the cat states defined above. It is worth noticing that, at variance with the emergence of the stationary cat-bistability at $\kappa_1 = 0$, where a classical bit can be stored, here a qubit can actually be stored, as long as it concerns the metastable time interval. Indeed, the coherences between the two cat states are preserved, too. We analyze such a quantum metastable regime by exploiting the spectral analysis of the Liouvillian, for a wide range of the model parameters within the regime of interest. The numerical analyses that follows are performed with a truncation at $n=25$ of the Hilbert space dimension, ensuring convergence of $10^{-5}$.

\begin{figure}
    \centering \includegraphics[width=\linewidth]{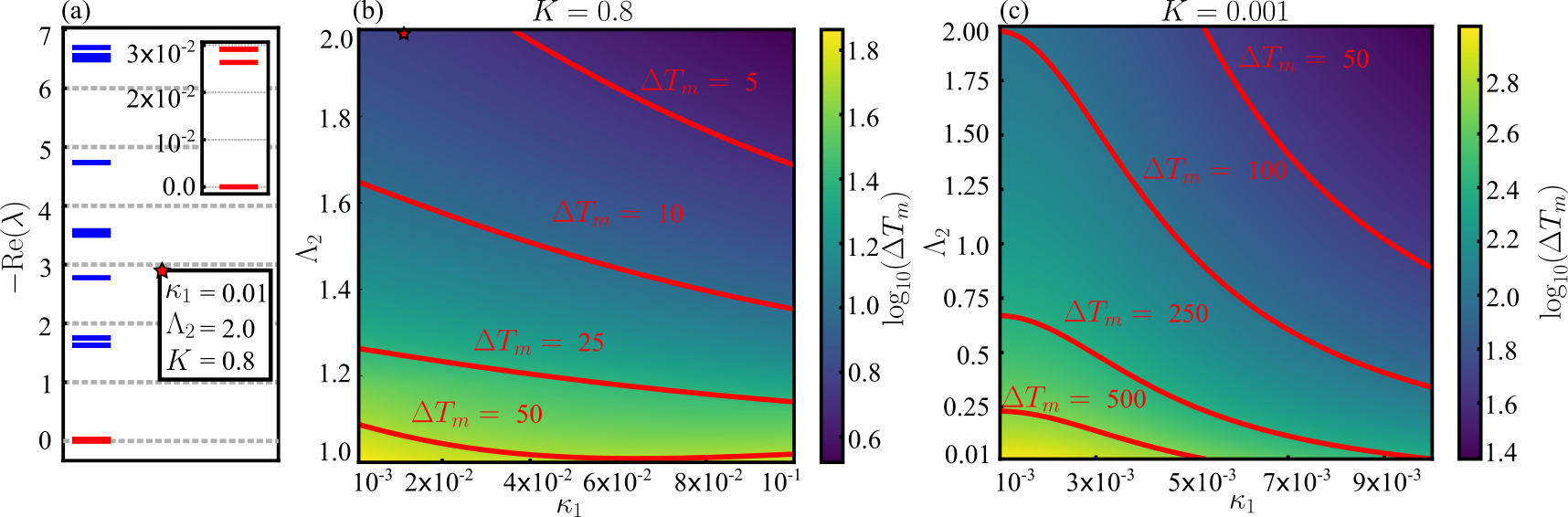}(b)
    \caption{Quantum metastability of  the two-photon driven-dissipative Kerr non-linear resonator subject to photon decay. We set $\kappa_2 = 1.$ Panel (a) shows the first $13$ elements in the spectrum of the Liouvillian, in terms of their real part, the inset considering the first four eigenvalues -- two of them with the same real part. We set $\Lambda_2 = 2.0$, $\kappa_1 = 0.01$, $K=1$. (b) Metastable time interval $\Delta T_m$ upon varying the single-photon loss rate $k_1$ and the two-photon driving $\Lambda_2$, with $K=0.8$. The (red) lines highlight from the the right-upper corner to the left-bottom one the following metastable time interval $ \Delta T_m = \left\lbrace 5, 10, 25, 50 \right\rbrace $. (b) We set $K=0.001$ and display the metastable time window as the two-photon driving $\Lambda_2$ and the single-photon decay rate $\kappa_1$ are varied, with the red, continuous lines representing $\Delta T_m = \left\lbrace 50, 100, 250, 500\right\rbrace$.}
    \label{fig:NLKerr_metastability_Lambda2}
\end{figure}

As already pointed out, we inspect the metastable regime by considering the gap between two consecutive eigenvalues of the Liouvillian and the rest of the spectrum, as its presence gives rise to a non-negligible metastable transient time $\Delta T_m$. The results are illustrated in Fig.~\ref{fig:NLKerr_metastability_Lambda2}. Panel (a) shows the first $13$ eigenvalues of the Liouvillian, in terms of their decreasing real part, while the inset displays the first four eigenvalues, highlighting the gap between the $4-$th and $5-$th eigenvalues, for the choice $K=0.8$, $\kappa_1 = 0.01$ $\kappa_2=1$, $\Lambda_2 = 2.0$. Panels (b) and (c) characterize metastability time windows by varying the single-photon decay, $\kappa_1$, and the two-photon driving, $\Lambda_2$, while fixing $k_2= 1$, and $K=0.08$ (b), $K=0.001$ (c). As we can see, the metastable transient regime becomes shorter while increasing both the single-photon loss rate $\kappa_1$, as well as the two-photon driving $\Lambda_2$. Moreover, it also decreases as the value of the Kerr interaction $K$ increases. The latter occurrence can also be appreciated by comparing the results of the metastable time interval in (b) and (c) of the same Fig.~\ref{fig:NLKerr_metastability_Lambda2}. 

\begin{figure}
    \centering
    \includegraphics[width=1\linewidth]{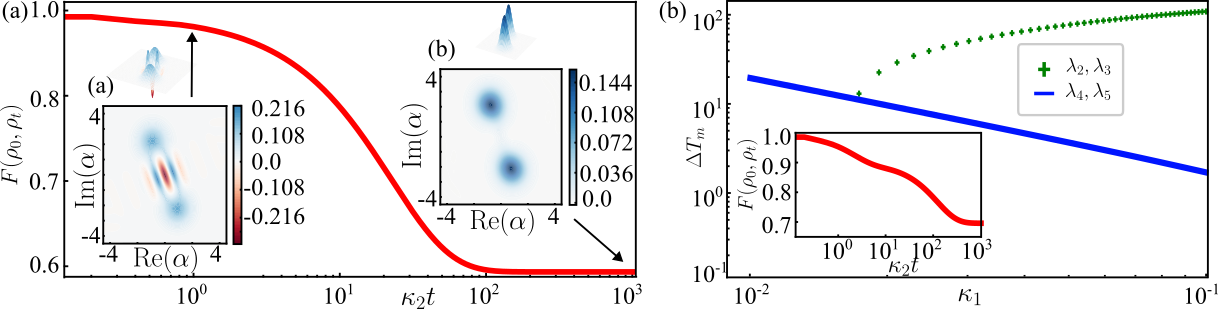}
    \caption{(a) Evolution through quantum metastability. Fidelity between the initial state, $\rho_0$, and the evolved state, $\rho_t$, as a function of time. We fix $\kappa_2=1$, $K=1$, $\kappa_1=0.01$, $\Lambda_2=4$. The initial state is chosen as a state belonging to the metastable DFS, $\ket{\psi_0} = \ket{C_{\alpha,-}}$. The inset, from left to right, show the Wigner representation of the state evolved at the time $\kappa_2 t =1$, and $\kappa_2 t =800$, respectively. (b) Crossing from quantum to classical metastability. Time interval $\Delta T_m$ during which eigenmodes of the Liouvillian associated to the $\lambda_i$-th  eigenvalue decay in function of the single-photon rate $\kappa_1$. Results are displayed in a logarithmic scale. Blue, thick line corresponds to the metastable DFS spanned by the cat-states, and it originates from the gap between the $5$-th and $4$-th eigenvalues of the Liouvillian. The green points (+ marker) represent instead the typical decay time of the third eigenmode of the Liouvillian. The values of the fixed parameters are $\Lambda_2 = 4$, $\kappa_2=1$, $K =1$. The inset represents the fidelity between the initial state and the evolved state vs time $t$ at $\kappa_1 = 0.1$.} 
    \label{fig:boson_fidelity time}
\end{figure}

An instance of the evolution through metastability is given in Fig.~\ref{fig:boson_fidelity time}(a). Here, we set the initial state inside the metastable DFS, specifically, we choose $\ket{\psi(0)} = \ket{C_{\alpha,-}}$, the odd superposition of the coherent states $\ket{\pm \alpha}$. We let the dynamics evolve under the action of the Hamiltonian and jump operators as defined by Eq.~\eqref{eq:lindblad_kerr}. The dynamics remains within the DFS during a metastable transient time $\Delta T_m \approx 5$, with finite coherences between the cat states, as highlighted in the Wigner representation of the state represented in the inset in the bottom left corner. Eventually, the system reaches the unique stationary state, which is a mixture of the two coherent states, as highlighted in the inset in the upper right corner.
In addition, we investigate how the quantum metastable regime illustrated above disappears when increasing the single-photon loss rate $\kappa_1$, to make it comparable to the two-photon decay $\kappa_2$. As a result, the coherences decay faster, up to the point where there is no trace left of metastability involving the whole manifold. Rather, a $2$-dimensional and classical metastability emerges: the two cat states are indeed metastable, although no coherence between them can survive. The regime of quantum metastability and the crossing towards the classical one is shown in Fig.~\ref{fig:boson_fidelity time}(b). We report the metastable time regime due to the separation between the $4-$th and $5-$th eigenvalues (red line), characterizing the metastable DFS, and the metastable transient regime due to the separation between the $2-$nd and $3-$rd eigenvalues (green marker +), possibly responsible for the emergence of classical metastability. The transition between the quantum metastability and the classical one can be seen taking place in a parameter regime where the single-photon decay rate is kept $\kappa_1 > 10^{-2}$. When increasing $\kappa_1$, the gap between $\lambda_5$ and $\lambda_4$ decreases, resulting in a shorter metastable transient time. Moreover, there exists a value $\kappa_1$ for which the gap between $\lambda_2 $ and $\lambda_3$ takes a finite value, inducing a classical metastability: i.e. a time interval, represented by the green dashed line (marker +) where the two cat states are metastable. An example of time evolution in this case is represented in panel (b), where we display the time evolution of the fidelity between the evolved state and the initial one, this chosen as a state belonging to the DFS spanned by the cat state.

\subsection{Error recovery protocol for a dephasing channel}

Before going ahead, it is worth noticing that when focusing on the parameter regime featuring quantum metastability, during the metastable transient itself, the system is protected from the leakage out of the code space induced by the single-photon decay. In the following, while considering the model in the parameter regime exhibiting metastable DFS, we will deal with possible errors induced by dephasing. We will therefore consider the following jump operator,
\begin{equation}
    J_{\phi} = \sqrt{k_{\phi}} a^{\dagger} a \, ,
\end{equation}
and the corresponding Liouvillian superoperator,
\begin{equation}\label{eq:liouvillian_deph_bos}
    \mathcal{L}'[\cdot] = {k_{\phi}} D_{a^{\dagger} a} (\cdot) \, ,
\end{equation}
Similarly to the previous section, in order to protect the system with respect to dephasing, we thus consider the following protocol [see Fig.~\ref{fig: 2qubitserrors}(a)]:
\begin{itemize}
    \item[$1)$] Initialize the system in a state belonging to the DFS metastable manifold of the unperturbed Liouvillian $\mathcal{L}_0$, e.g. $\rho_0 = \ket{\psi}\bra{{\psi}}$, where $\ket{\psi} = c_+\ket{C_{\alpha,+}} + c_- \ket{C_{\alpha,-}} $.
    \item[$2)$] Let an error occur in terms of the additional Liouvillian $\mathcal{L'}$ defined by Eq.~\eqref{eq:liouvillian_deph_bos}, which acts for a time $\Delta t_e << \Delta T_{\mathrm{m}}$, so that the initial state evolves into $\rho(\Delta t_e) = (\mathcal{L}_0 +\mathcal{L}')[\rho_0]$.
    \item[$3)$] Let the dynamics ruled by the unperturbed Liouvillian $\mathcal{L}_0$, defined by Eq.~\eqref{eq:lindblad_kerr}, that evolves the state $\rho(\Delta t_e)$ up to a final time, say $t_f$, $\rho(t_f) = \mathcal{L}_0[\rho(\Delta t_e)]$.
\end{itemize}
We characterize the protocol described above through the fidelity, by analyzing to what extent the final state resembles the initial one, $\rho(t_f) \approx  \rho_0$ , and, in addition, we characterize the time needed to recover the state when the fidelity reaches its maximum value. 

\begin{figure}
    \centering    \includegraphics[width=1\linewidth]{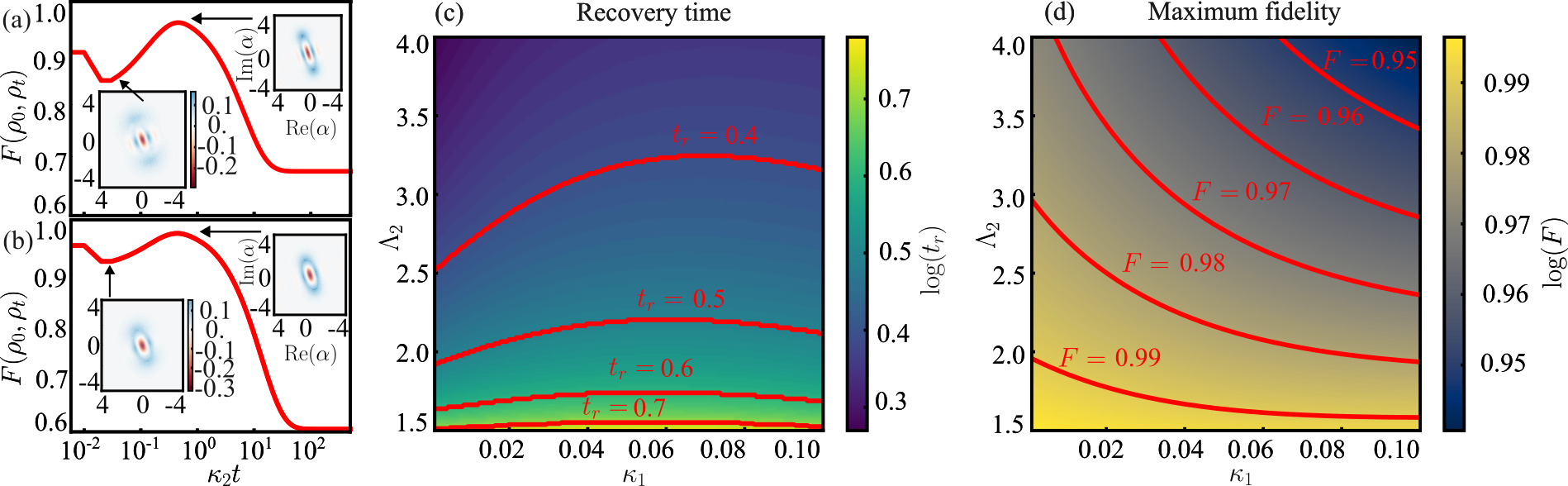}
    \caption{Error recovery performance. We set $\kappa_{\phi} = 10$, $K=1$, $\kappa_2 = 1$. (a), (b) Fidelity between the initial state, taken as $\ket{\psi_0} = \ket{C_{\alpha,-}}$, and the state  $\rho_t$ evolved at time $t$. The evolution is made through the Liouvillian $\mathcal{L}_0 + \mathcal{L'}$ in the time interval $[0, t_e ]$, with $k_2 t_e = 0.03$ in the figure. After the time $t_e$, the system is evolved by means of the Liouvillian $\mathcal{L}_0$, up to a final time $t_f$, with $\kappa_2 t_f=500$. We further set $\kappa_1 = 0.04$, and (a)  $\Lambda_2=3.5$, (b) $\Lambda_2=2.0$. (c) Recovery time $\kappa_2 t_r$ of the initial state upon varying $\kappa_1$, $\Lambda_2$. The recovery time corresponds to the time at which the fidelity between the evolved state and the initial one takes the maximum value after the error occurred. (d) Value of fidelity at recovery time $\kappa_2 t_r$ upon varying $\kappa_1$, $\Lambda_2$.}
    \label{fig:error_protocol boson}
\end{figure}

The results are summarized in Fig.~\ref{fig:error_protocol boson}. As already specified, the main quantity we consider is the fidelity $F(\rho_0, \rho_t)$, between the initial state $\rho_0$, and the evolved state $\rho_t$. In panels (a), (b), we show the fidelity as a function of time, when the above-defined protocol is performed. Consistently with step $1)$, we take an initial state belonging to the metastable DFS of the Liouvillian $\mathcal{L}_0$, and specifically $\ket{\psi_0} = \ket{C_{\alpha,-}}$, $\alpha$ being fully specified by the parameter values chosen and reported in the figure caption. Following step $2)$, the state is subject to an error, and thus is evolved by means of the Liouvillian $\mathcal{L}_0 + \mathcal{L}'$, up to a time $\kappa_2 t_e$. After such a time, the dephasing channel is turned off and the state is evolved through the Liouvillian $\mathcal{L}_0$ only, as described in step $3)$. Here, the unperturbed Lindblad evolution $\mathcal{L}_0$ is indeed capable of evolving the state of the system back to the very same initial state inside the DFS manifold, up to a certain fidelity value, as shown in the figure. Panels (a) and (b) are further characterized by $\Lambda_2 = 3.5$ and $\Lambda_2 = 2$, respectively, and we can see that a different state is reached at time $\kappa_2 t_e$. Similarly, it can be illustrated that a different faulty state is reached upon changing the error strength in terms of the rate $\kappa_{\phi}$. Instead, when the error channel is turned off, we characterize the recovery by means of the time needed to reach the state that features the maximum value of the fidelity with respect to the initial one. We refer to this time as the recovery time $t_r$ measured --as in the other cases-- in terms of the rate $\kappa_2$. Panel (c) displays a color map of the recovery time $\kappa_2 t_r$ as a function of the single-photon loss rate $\kappa_1$ and the two-photon driving parameter $\Lambda_2$. The protocol is the same as the one employed for panels (a), (b), with $\ket{\psi_0} = \ket{C_{\alpha,-}}$, $\kappa_{\phi} = 10$. The recovery time becomes shorter as the parameter $\Lambda_2$ increases. It is worth stressing that $t_r$ fully depends on the elements of the Liouvillian spectrum $\lambda_k$, with $k\geq 5$. However, evaluating the contribution of each element, which depends on the overlap of the given state at time $t_e$ with the right eigenmatrices $R_k$, would require to have access to the state $\rho(t_e)$.  Panel (d) shows the maximum value of the fidelity after the error channel has been switched off, upon varying the single-photon loss rate $\kappa_1$ and the two-photon driving parameter $\Lambda_2$. It is obtained as the fidelity between the initial state and the one evolved at the recovery time $\kappa_2 t_r$. Upon comparing panel (c), (d), it is worth highlighting that a shorter recovery time can be obtained at the cost of having a lower fidelity, and vice versa.

\section{Discussion}
In this work, we approach the paradigm of passive error correction leveraging the intrinsic dynamics of open quantum systems to autonomously recover from certain types of errors. Specifically, we focus on code spaces defined within quantum metastable manifolds, where DFS persist over long transient times before the system relaxes to its unique steady state. In this scenario, we introduced a protocol in which an initial state is prepared within a metastable DFS, subject to errors that can occur for a short time, and then the state is left to evolve under the unperturbed dynamics. Here, our goal was to assess whether and to what extent the unperturbed dynamics can map the system back to the initial state. More concretely, we focus first on a model represented by two qubits, which are subject to collective dissipation, the metastable DFS being spanned by the states $\ket{01}$, $\ket{10}$. We deal with different types of errors, bit-flip errors, spontaneous emission, and phase-flip errors. The first two types of errors take the system out of the DFS and can be almost effectively corrected by the metastable dynamics, restoring the initial state with high fidelity. Phase-flip errors, in contrast, keep the system approximately within the DFS but suppress coherences, making impossible the full recovery through the passive dynamics of the unperturbed Liouvillian. In all the cases, there happens to be a relatively small but inevitable leakage out of the code space, which could potentially cause non-negligible issues when attempting to further manipulate the state within the code space. We then consider a bosonic Kerr resonator with two-photon driving and dissipation, in the parameter regime that supports a metastable DFS spanned by cat states. We examined the effect of a dephasing channel and found that, across a broad range of parameters, the metastable dynamics can partially recover the corrupted information. However, we remark a trade-off: faster recovery comes at the cost of lower fidelity. 

These findings highlight both the potential and limitations of using metastable code spaces for quantum information processing. On the one hand, they are unsuitable for long-term quantum memory, as the system eventually relaxes to a typically trivial steady state. On the other hand, however, metastable manifolds offer a promising avenue for application in quantum protocols, where quantum states are used soon after preparation within a circuit or algorithm. Indeed, despite their imperfections, they could be employed to design more robust physical qubits. This study lays the ground for new passive error mitigation strategies based on metastable dynamics. Future work will focus on extending the approach to explore its issues and robustness under quantum gates \cite{MirrahimiEtAl14}, and investigating alternative characterizations such as unravelings of the master equation \cite{Brown2016Nov}.

\section*{Acknowledgment}
The authors thank Hugo Perrin for nice and insightful discussions. EF acknowledges funding from the European Union’s Horizon Europe program through Grant No. 101105267, and the Spanish State Research Agency, through the Maria de Maeztu project CEX2021-001164-M. TB acknowledges funding from the CNRS through the EMERGENCE@INC2024 project DINOPARC, and the French National Research Agency under the Investments of the Future Program project ANR-21-ESRE-0032 (aQCess).

\section*{Statement}
The authors contributed equally to this work.

\bibliography{references}

\end{document}